\documentclass[preprint,showpacs,preprintnumbers,amsmath,amssymb]{revtex4}

\usepackage{amssymb}
\usepackage{amsmath}
\usepackage{graphicx}
\usepackage{epsfig}
\usepackage{subfigure}
\usepackage{amsfonts}
\usepackage{CJK}

\begin{document}

\title{Circuit QED: Implementation of the three-qubit refined Deutsch-Jozsa quantum algorithm}

\author{Qi-Ping Su and Chui-Ping Yang}

\address{Department of Physics, Hangzhou Normal University,
Hangzhou, Zhejiang 310036, China}
\date{\today}

\begin{abstract}
We propose a protocol to construct the 35 $f$-controlled phase gates of a three-qubit
refined Deutsch-Jozsa (DJ) algorithm, by using single-qubit $\sigma_z$ gates,
two-qubit controlled phase gates, and two-target-qubit controlled phase gates.
Using this protocol, we discuss how to implement the three-qubit refined DJ
algorithm with superconducting transmon qutrits resonantly coupled to a single cavity.
Our numerical calculation shows that implementation of this quantum
algorithm is feasible within the present circuit QED technique. The experimental
realization of this algorithm would be an important step toward more complex quantum
computation in circuit QED.
\end{abstract}

 \maketitle
\date{\today}

\section{INTRODUCTION}

As one of the most promising solid-state candidates for quantum information
processing [1,2], the physical system composed of circuit cavities and
superconducting qubits is of particular interest. This is because: (i)
superconducting qubits and microwave resonators (a.k.a. cavities) can be
fabricated using modern integrated circuit technology, and their properties
can be characterized and adjusted in situ, (ii) superconducting qubits have
relatively long decoherence times [3,4], and (iii) a superconducting
microwave cavity or resonator plays a role of quantum bus\ which can mediate
long-range and fast interaction between distant superconducting qubits
[5-7]. Moreover, the strong coupling between the cavity field and
superconducting qubits, which is difficult to implement with atoms in a
microwave cavity, was earlier predicted in theory [1,8] and has been
experimentally demonstrated [9,10]. Because of these features, circuit QED
has been widely utilized for quantum information processing. Based on
circuit QED, many theoretical proposals have been presented for realizing
two-qubit gates [5,11-18] and multiple qubit gates [19-20] with
superconducting qubits. Moreover, experimental demonstration of two-qubit
gates [6,7,21,22] and three-qubit gates [23-25] with superconducting qubits
in circuit QED has been also reported.

The interest in quantum computation is stimulated by the discovery of
quantum algorithms [26,27] which can solve problems of significance much
more efficiently than their classical counterparts. Among important quantum
algorithms, there exist the Deutsch algorithm [28], the Deutsch-Jozsa
algorithm [29], the Shor algorithm [30], the Simon algorithm [31], the
quantum Fourier transform algorithm, and the Grover search algorithm [32].
As is well known, the Deutsch algorithm and the Deutsch-Jozsa algorithm were
the first two that make use of the features of quantum mechanics for quantum
computation. Compared with other quantum algorithms, these two algorithms
are easy to be implemented and thus have been considered as the natural
candidates for demonstrating power of quantum computation.

We note that with superconducting qubits coupled to a circuit cavity, a
\textit{two-qubit} \textit{Deutsch-Jozsa} quantum algorithm and a two-qubit
Grover search quantum algorithm were previously demonstrated in experiments
[7]. However, after a thorough investigation, we note that how to implement
a \textit{three-qubit} Deutsch-Jozsa (DJ) quantum algorithm with
superconducting qubits or qutrits in circuit QED has not been reported in
both theoretical and experimental aspects.~As is known, the experimental
realization of the three-qubit DJ algorithm with a
cavity-superconducting-device system is important because it would be an
important step toward more complex quantum computation in circuit QED.

In this paper, we propose a protocol to construct the 35 $f$-controlled
phase gates of a three-qubit refined DJ algorithm, by using single-qubit $%
\sigma _z$ gates, two-qubit CP gates, and two-target-qubit CP gates. It
should be noted that a two-target-qubit CP gate consists of two sequential
controlled $\sigma _z$ gates, which have a common control qubit but a
different target qubit. The protocol is quite general and can be applied to
implement the three-qubit DJ algorithm in various of physical systems. Based
on this protocol, we further discuss how to implement the three-qubit
refined DJ algorithm with superconducting transmon qutrits resonantly
coupled to a single cavity. Our numerical calculation shows that
implementation of this quantum algorithm is feasible within the present
circuit QED technique.

The paper is organized as follows. In Sec.~II, we review the refined DJ
algorithm. In Sec. III, we present a protocol to construct the 35 $f$%
-controlled phase gates. In Sec.~IV, we discuss how to implement this DJ
algorithm with superconducting transmon qutrits coupled to a cavity via
resonant interaction, and analyze the experimental feasibility. A concluding
summary is given in Sec.~V.

\section{REFINED DEUTSCH-JOZSA ALGORITHM}

The DJ algorithm is aimed at distinguishing the constant function from the
balanced functions on $2^{n}$ inputs. The function $f\left( x\right) $ takes
either $0$ or $1.$ For the constant function, the function values are
constant ($0$ or $1$) for all $2^{n}$ inputs. In contrast, for the balanced
function, the function values are equal to 1 for half of all the possible
inputs, and $0$ for the other half. Using the DJ algorithm, whether the
function is constant or balanced can be determined by only one query.
However, a classical algorithm would require $2^{n-1}+1$ queries to answer
the same problem, which grows exponentially with input size.

\begin{figure}[tbp]
\includegraphics[bb=134 551 469 642, width=12.5 cm, clip]{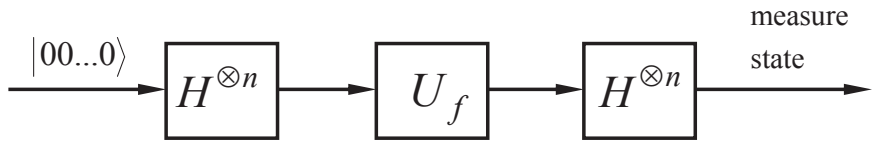} \vspace*{%
-0.08in}
\caption{Quantum circuit for the refined $n$-qubit Deutsch-Jozsa algorithm.}
\label{fig:1}
\end{figure}

The refined DJ algorithm was proposed by Collins \textit{et al.} in 2001
[33], which is illustrated in Fig.~1. This refined DJ algorithm is described
below:

(i) Each input query qubit is prepared in the initial state $\left|
0\right\rangle .$

(ii) Perform a Hadamard transformation $H$ on each qubit, resulting in $%
\left\vert 0\right\rangle \rightarrow \left( \left\vert 0\right\rangle
+\left\vert 1\right\rangle \right) /\sqrt{2}$ and $\left\vert 1\right\rangle
\rightarrow \left( \left\vert 0\right\rangle -\left\vert 1\right\rangle
\right) /\sqrt{2}$). As a result, the $n$-qubit initial state $\left\vert
00\cdot \cdot \cdot 0\right\rangle $ changes to the state $\frac{1}{2^{n/2}}%
\sum_{x=0}^{2^{n}-1}\left\vert x\right\rangle $ (denoted as $\left\vert \psi
_{1}\right\rangle $)$.$

(iii) Apply the $f$-controlled phase shift $U_{f}$, described by
\begin{equation}
\left\vert x\right\rangle \overset{U_{f}}{\longrightarrow }\left( -1\right)
^{f\left( x\right) }\left\vert x\right\rangle ,
\end{equation}
which leads the state $\left\vert \psi _{1}\right\rangle $ to the state $%
\frac{1}{2^{n/2}}\sum_{x=0}^{2^{n}-1}\left( -1\right) ^{f\left( x\right)
}\left\vert x\right\rangle $ (denoted as $\left\vert \psi _{2}\right\rangle $%
).

(iv) Perform another Hadamard transformation $H$ on each qubit. As a result,
the state $\left\vert \psi _{2}\right\rangle $ becomes $\frac{1}{2^{n}}%
\sum_{z=0}^{2^{n}-1}\sum_{x=0}^{2^{n}-1}\left( -1\right) ^{x\cdot z+f\left(
x\right) }\left\vert z\right\rangle .$

(v) Measure the final state of the $n$ qubits. If the $n$ qubits are
measured in the state $\left\vert 00...0\right\rangle ,$ then $f\left(
x\right) $ is constant. However, if they are measured in other $n$-qubit
computational basis states, then $f\left( x\right) $ is balanced. This is
because the amplitude $a_{\left\vert 00...0\right\rangle }$ of the state $%
\left\vert 00...0\right\rangle $ is given by $a_{\left\vert
00...0\right\rangle }=\frac{1}{2^{n}}\sum_{x=0}^{2^{n}-1}\left( -1\right)
^{f\left( x\right) },$ which is $\pm 1$ for a constant $f\left( x\right) $
while $0$ for a balanced $f\left( x\right) $.

One can see that when compared with the original DJ algorithm [29], this
refined DJ algorithm does not need an auxiliary working qubit. Hence, it
requires one qubit fewer than the original DJ algorithm. Consequently, its
physical implementation requires one fewer two-state system.

\begin{table}[tbp]
\includegraphics[bb=0 0 800 600, width=13.5 cm, clip]{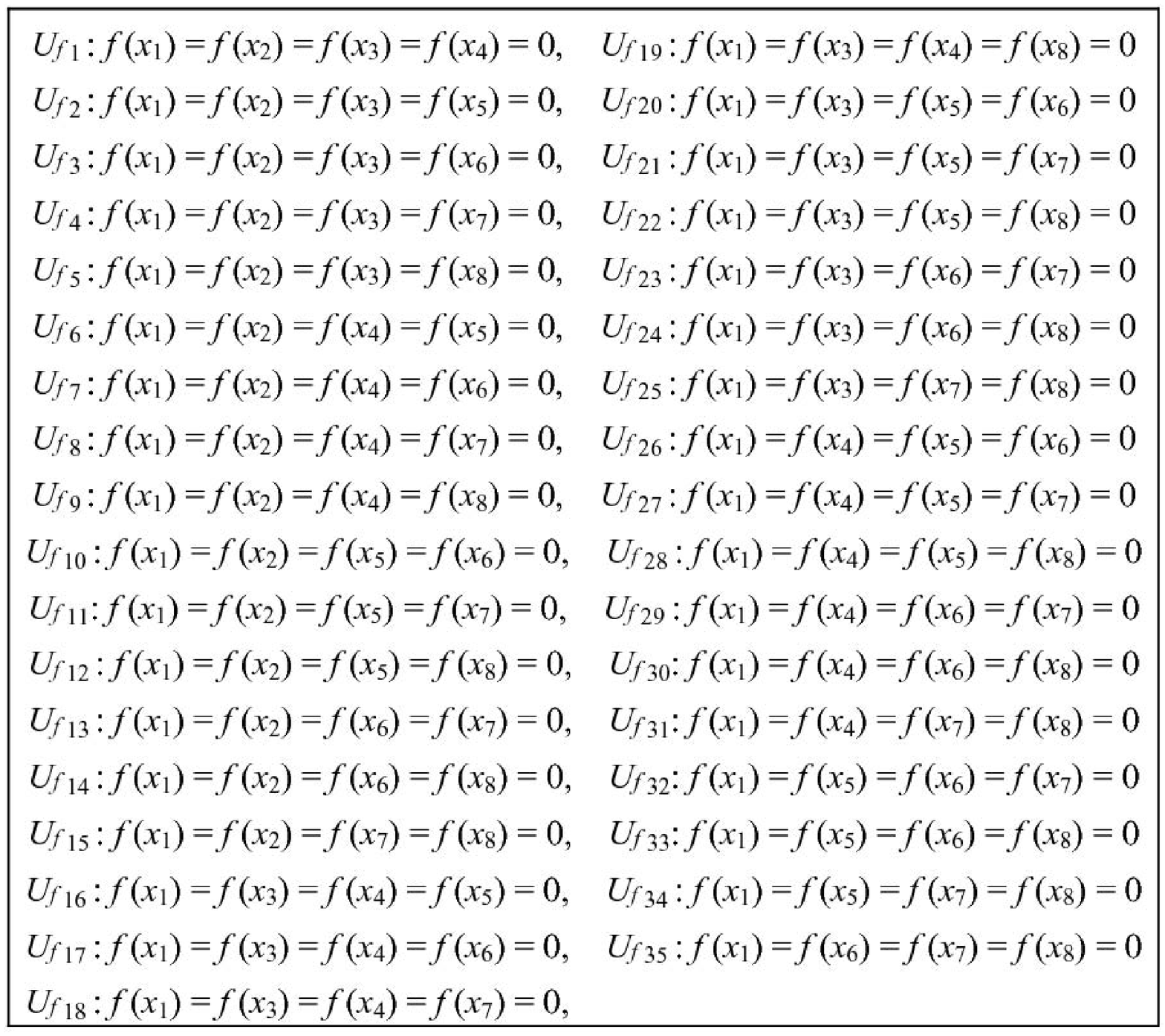} \vspace*{%
-0.08in}
\caption{List of the 35 balanced functions for a three-qubit refined
Deutsch-Jozsa quantum algorithm. Here, $%
x_{1}=000,x_{2}=001,x_{3}=010,x_{4}=011,x_{5}=100,x_{6}=101,x_{7}=110, $ and
$x_{8}=111.$ For simplicity, we only list the funtion values, which are $%
``0" $, for four inputs corresponding to each balanced function. Note that
for each balanced function, the function values for the other four inputs
(not listed) take a value $``1"$. For instance, for the balanced function
corresponding to $U_{f1},$ the function values for the other four inputs
(not listed) are $f\left( x_{5}\right) =f\left( x_{6}\right) =f\left(
x_{7}\right) =f\left( x_{8}\right) =1.$}
\label{table:1}
\end{table}

\begin{table}[tbp]
\includegraphics[bb=0 0 1200 900, width=13.5 cm, clip]{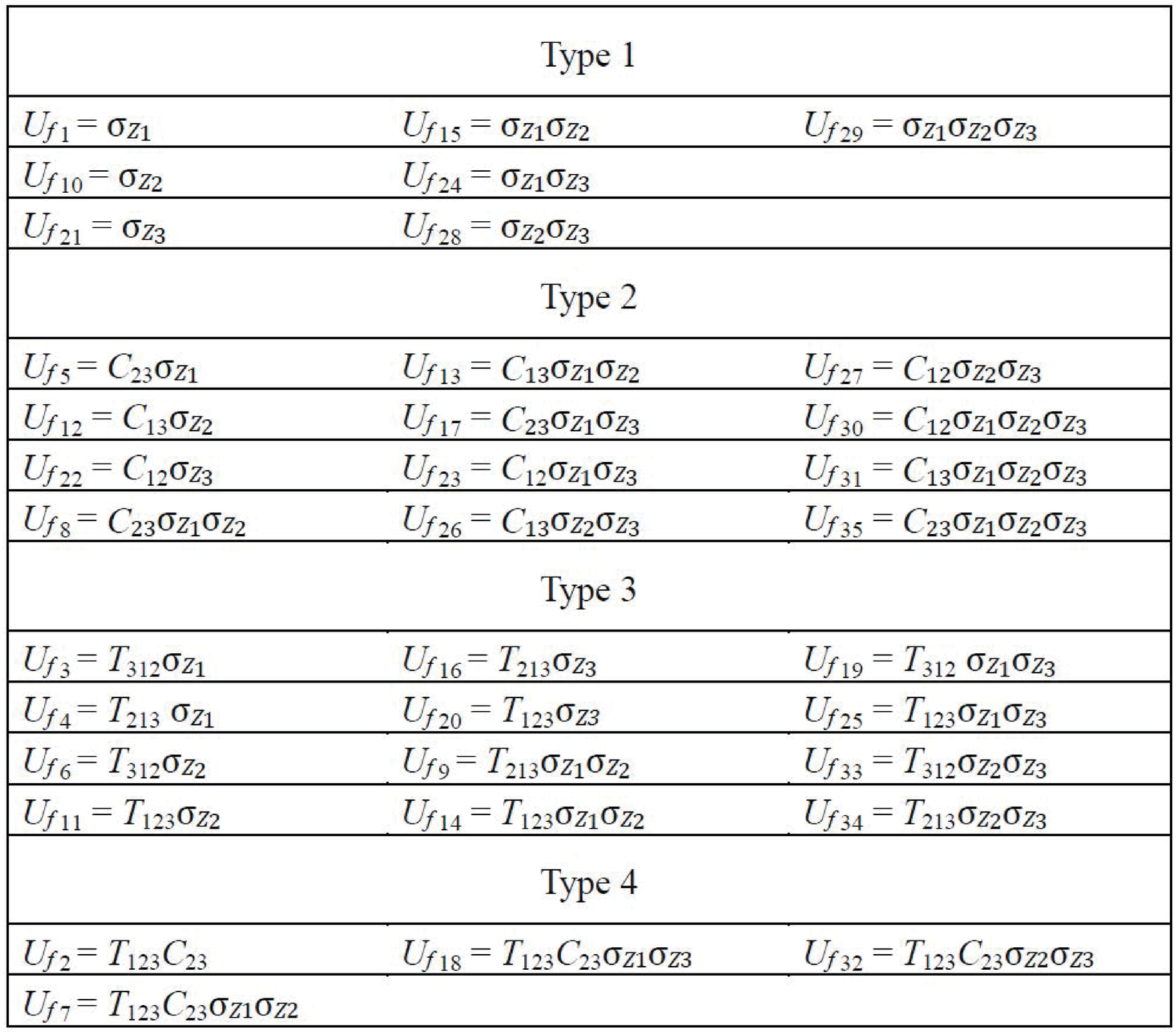} \vspace*{%
-0.08in}
\caption{List of 35 $f$-controlled phase gates for a three-qubit refined
Deutsch-Jozsa quantum algorithm. Here, $\protect\sigma_{z_j}$ represents a
single-qubit $\protect\sigma_{z_j}$ gate on qubit $j$ ($j=1,2,3$); $C_{jk}$
indicates a two-qubit controlled-phase gate on qutrits $j$ and $k$ ($%
j,k=1,2,3$), described by Eq.~(2); and $T_{jkl}$ is a two-target-qubit CP
gate described by Eq.~(3).}
\label{table:2}
\end{table}

\section{PROTOCOL FOR CONSTRUCTION OF THE $f$-CONTROLLED PHASE GATES}

For a $n$-qubit DJ algorithm, there are a total of $C_{2^{n}}^{2^{n-1}}$
balanced functions, among which only $C_{2^{n}}^{2^{n-1}}/2$ balanced
functions are nontrivial if the symmetry is taken into account. For the
three-qubit DJ algorithm, i.e., $n=3,$ there thus exist $C_{8}^{4}/2=35$
nontrivial balanced functions $U_{f1},U_{f2},...,U_{f35}$ (see Table I). In
this section, we show how to construct the 35 $f$-controlled phase gates, by
using single-qubit $\sigma _{z}$ gates, two-qubit CP gates, and
two-target-qubit CP gates.

A single qubit $\sigma _{z}$ gate results in the transformation $\sigma
_{z}\left\vert 0\right\rangle =\left\vert 0\right\rangle $ while $\sigma
_{z}\left\vert 1\right\rangle =-\left\vert 1\right\rangle .$ A two-qubit CP
gate $C_{jk}$ on qubits $j$ and $k$ considered here is described as follows
\begin{equation}
\left\vert mn\right\rangle _{jk}\rightarrow \left( -1\right) ^{m\times
n}\left\vert mn\right\rangle _{jk};\,\,m,n\in \left\{ 0,1\right\}
\end{equation}
which implies that if and only if the control qubit $j$ (the first qubit) is
in the state $\left\vert 1\right\rangle $ , a phase flip happens to the
state $\left\vert 1\right\rangle $ of the target qubit $k$ (the second
qubit), but nothing happens otherwise. In addition, a two-target-qubit CP
gate $T_{jkl}$ with the control qubit $j$ and the two target qubits $k$ and $%
l$ is defined below
\begin{equation}
\left\vert mnr\right\rangle _{jkl}\rightarrow \left( -1\right) ^{m\times
n}\left( -1\right) ^{m\times r}\left\vert mnr\right\rangle
_{jkl};\,\,m,n,r\in \left\{ 0,1\right\}
\end{equation}
which shows that if and only if the control qubit $j$ (the first qubit) is
in the state $\left\vert 1\right\rangle ,$ a phase flip happens to the state
$\left\vert 1\right\rangle $ of the target qubit $k$ (the second qubit) and
the state $\left\vert 1\right\rangle $ of the target qubit $l$ (the last
qubit).

The construction for each of the 35 $f$-controlled phase gates is listed in
Table II. One can see from Table II that the 35 $f$-controlled phase gates $%
U_{f1},U_{f2},...,U_{f35}$ are classified into the following four types: (i)
Type 1 includes seven $f$-controlled phase gates each constructed with
single-qubit $\sigma _{z}$ gates only; (ii) Type 2 contains twelve\ $f$%
-controlled phase gates each constructed with single-qubit $\sigma _{z}$
gates and one two-qubit CP gate; (iii) Type 3 has twelve $f$-controlled
phase gates each constructed by using single-qubit $\sigma _{z}$ gates and a
two-target-qubit CP gate; and (iv)\ Type 4 involves four $f$-controlled
phase gates each implemented with single-qubit $\sigma _{z}$ gates, a
two-qubit CP gate, and a two-target-qubit CP gate at most.

\section{IMPLEMENTATION OF THE THREE-QUBIT REFINED DJ ALGORITHM IN CIRCUIT
QED}

Using the protocol, we now discuss how to implement the three-qubit refined
DJ algorithm with three superconducting transmon qutrits ($1,2,3$) each
having three levels (i.e., the ground $\left| 0\right\rangle ,$ the first
excited $\left| 1\right\rangle ,$ and the second excited level $\left|
2\right\rangle $). We then estimate the fidelity of the operation, which is
performed in a setup composed of three phase qutrits and a one-dimensional
coplanar waveguide resonator.

\subsection{Implementing the algorithm}

Since the $f$-controlled phase gates belonging to Type 1 are constructed
using single-qubit gates only, their implementation does not require
entanglement and thus can be realized in a classical way. Therefore, in the
following we focus on the $f$-controlled phase gates belonging to Type 2,
Type 3, and Type 4. Without loss of generality, let us consider the three $f$%
-controlled phase gates $U_{f30}$ (belonging to Type 2)$,$ $U_{f9\text{ }}$%
(belonging to Type 3)$,$ and $U_{f7}$ (belonging to Type 4). By comparing
them with other $f$-controlled phase gates in the same types, it can be
found that these three unitary gates $U_{f30},U_{f9},$ and $U_{f7}$ contain
the same number of two-qubit CP gates or/and two-target-qubit CP gates but a
greater or equal number of single-qubit gates than the other $f$-controlled
phase gates in the same types. Hence, if the three $f$-controlled phase
gates $U_{f30}$, $U_{f9}$, and $U_{f7}$ can be implemented, then other $f$%
-controlled phase gates in the same types can be achieved with a higher or
equal fidelity. In this sense, to see how well the proposal works, it would
be sufficient to explore the implementation feasibility of the following
three joint quantum operations, described by
\begin{eqnarray}
U_1 &=&H^{\otimes 3}U_{f30}H^{\otimes 3}=H^{\otimes 3}C_{12}\sigma
_{z1}\sigma _{z2}\sigma _{z3}H^{\otimes 3},  \notag \\
U_2 &=&H^{\otimes 3}U_{f9}H^{\otimes 3}=H^{\otimes 3}T_{213}\sigma
_{z1}\sigma _{z2}H^{\otimes 3},  \notag \\
U_3 &=&H^{\otimes 3}U_{f7}H^{\otimes 3}=H^{\otimes 3}T_{123}C_{23}\sigma
_{z1}\sigma _{z2}H^{\otimes 3},
\end{eqnarray}
where the gate operation sequence is from right to left. The $U_1$, $U_2$,
and $U_3$ here are constructed, according to Fig. 1 and the decomposition of
$U_{f30}$, $U_{f9}$, and $U_{f7}$ given in Table II. From Eq.~(4), one can
see that $U_1,U_2,$ and $U_3$ are implemented through the single-qubit $%
\sigma _z$ and $H$ gates, two-qubit CP gates and two-target-qubit CP gates.

\textit{A.1 Implementing single-qubit gates---}The single-qubit Hamardard $H$
gate or $\sigma _z$ gate on qutrit $j$ can be realized by applying a pulse
resonant with the $\left| 0\right\rangle \leftrightarrow \left|
1\right\rangle $ transition of qutrit $j$ ($j=1,2,3$). To eliminate the
leakage into the level $\left| 2\right\rangle ,$ one can employ the DRAG
pulse [34,35], which can reduce the gate error by an order of magnitude
relative to the state of the art, all based on smooth and feasible pulse
shapes [34]. In addition, to shorten the gate time, the three\ joint
Hamardard gates $H^{\otimes 3}$ in Eq. (4) are performed simultaneously,
which can be achieved by turning on and off the pulses applied to the three
qutrits at the same time. In the same manner, the three $\sigma _z$ gates
involved in $U_1$ (the two $\sigma _z$ gates in $U_2$ and $U_3$) are
performed simultaneously.

Detailed discussion of how to implement the $H$ gate or $\sigma _z$ gate is
omitted here since implementing a single-qubit gate depends on the use of
the pulse shapes and is straightforward in experiments.
\begin{figure}[tbp]
\includegraphics[bb=91 431 522 663, width=10.5 cm, clip]{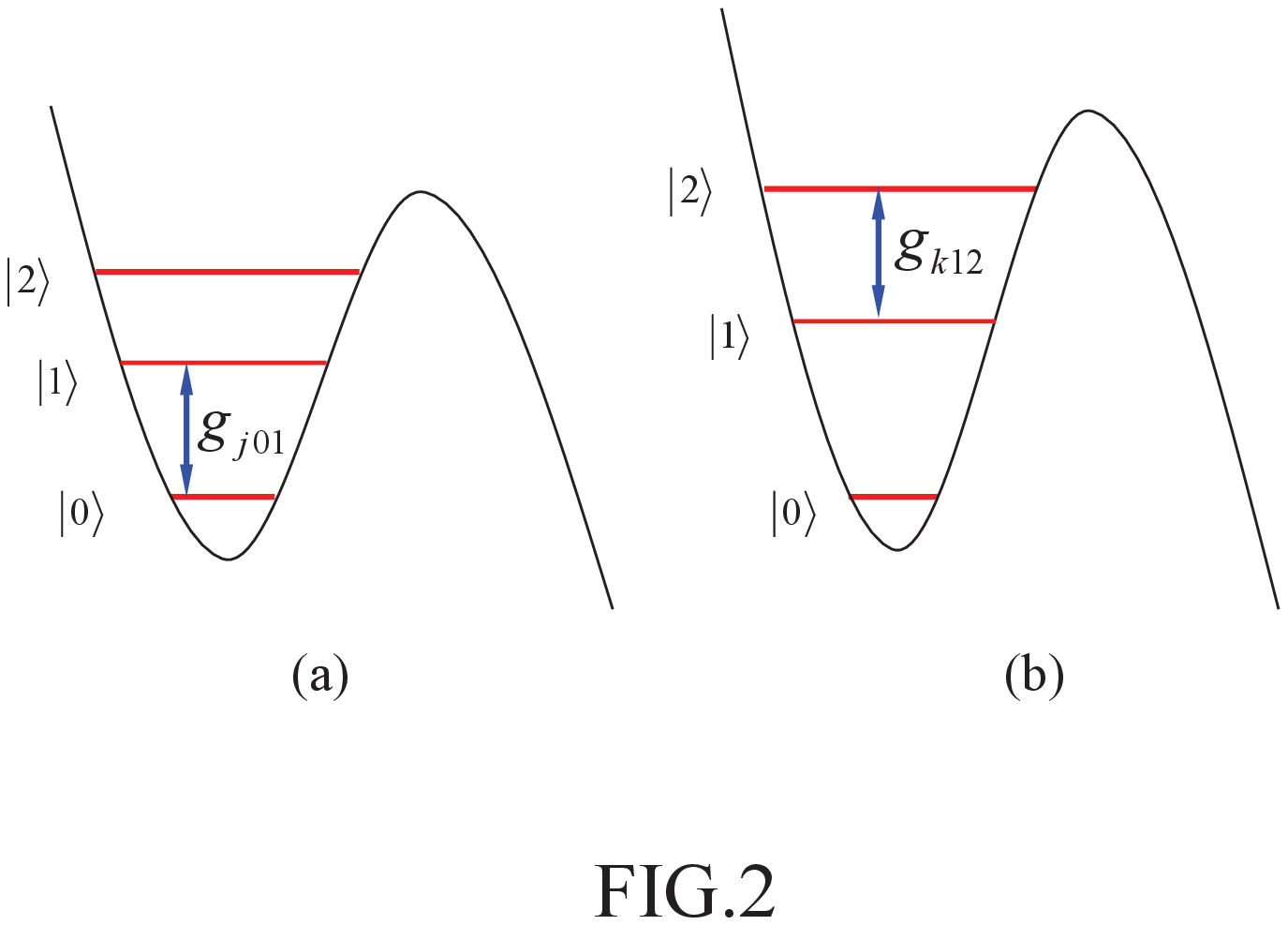} \vspace*{%
-0.08in}
\caption{(Color online) Illustration of qutrit-cavity resonant interaction.
(a) The cavity is resonant with the $\left| 0\right\rangle \leftrightarrow
\left| 1\right\rangle $ transition of qutrit $j$ with a coupling constant $%
g_{j01}$. (b) The cavity is resonant with the $\left| 1\right\rangle
\leftrightarrow \left| 2\right\rangle $ transition of qutrit $k$ with a
coupling constant $g_{k12}$.}
\end{figure}

\textit{A.2 Implementing a two-qubit CP gate--- }We define $g_{j01}$ ($%
g_{k12}$) as the resonant coupling constant between the cavity mode and the $%
\left\vert 0\right\rangle \leftrightarrow \left\vert 1\right\rangle $ ($%
\left\vert 1\right\rangle \leftrightarrow \left\vert 2\right\rangle $)
transition of qutrit $j$ ($k$). The cavity is initially in the vacuum state $%
\left\vert 0\right\rangle _{c}.$ The procedure for realizing $C_{jk}$ is
listed as follows:

Step (i). Adjust the level spacings of qutrit $j$ such that the $\left\vert
0\right\rangle \leftrightarrow \left\vert 1\right\rangle $ transition is on
resonance with the cavity [Fig.~2(a)]. After an interaction time $t_{1}=\pi
/\left( 2g_{j01}\right) $, the state $\left\vert 1\right\rangle
_{j}\left\vert 0\right\rangle _{c}$ changes to $-i\left\vert 0\right\rangle
_{j}\left\vert 1\right\rangle _{c}$ while nothing happens to the state $%
\left\vert 0\right\rangle _{j}\left\vert 0\right\rangle _{c}$ (e.g., see
[19]).

Step (ii). Adjust the level spacings of qutrit $k$ such that the $\left\vert
1\right\rangle \leftrightarrow \left\vert 2\right\rangle $ transition is on
resonance with the cavity [Fig.~2(b)]. After an interaction time $t_{2}=\pi
/g_{k12},$ the state $\left\vert 1\right\rangle _{k}\left\vert
1\right\rangle _{c}$ becomes $-\left\vert 1\right\rangle _{k}\left\vert
1\right\rangle _{c}$ while the states $\left\vert 0\right\rangle
_{k}\left\vert 0\right\rangle _{c},$ $\left\vert 1\right\rangle
_{k}\left\vert 0\right\rangle _{c}$ and $\left\vert 0\right\rangle
_{k}\left\vert 1\right\rangle _{c}$ remain unchanged.

Step (iii). Adjust the level spacings of qutrit $j$ such that the $%
\left\vert 0\right\rangle \leftrightarrow \left\vert 1\right\rangle $
transition is on resonance with the cavity [Fig.~2(a)]. After an interaction
time $t_{3}=3\pi /\left( 2g_{j01}\right) $, the state $\left\vert
0\right\rangle _{j}\left\vert 1\right\rangle _{c}$ changes to $i\left\vert
1\right\rangle _{j}\left\vert 0\right\rangle _{c}$ while nothing happens to
the state $\left\vert 0\right\rangle _{j}\left\vert 0\right\rangle _{c}.$

One can check that the state $\left\vert 1\right\rangle _{j}\left\vert
0\right\rangle _{k}\left\vert 0\right\rangle _{c}$ remains unchanged while
the state $\left\vert 1\right\rangle _{j}\left\vert 1\right\rangle
_{k}\left\vert 0\right\rangle _{c}$ changes to $-\left\vert 1\right\rangle
_{j}\left\vert 1\right\rangle _{k}\left\vert 0\right\rangle _{c}$ after the
above operations. On the other hand, the initial states \{$\left\vert
0\right\rangle _{j}\left\vert 0\right\rangle _{k}\left\vert 0\right\rangle
_{c},\left\vert 0\right\rangle _{j}\left\vert 1\right\rangle _{k}\left\vert
0\right\rangle _{c}$\} remain unchanged during the entire operation above.
These results show that a two-qubit CP gate $C_{jk},$ described by Eq.~(2),
was achieved with qutrit $j$ (the control) and qutrit $k$ (the target) after
the above process, while the cavity returns to its original vacuum state.

\textit{A.3 Realizing a two-target-qubit CP gate--- }By carefully examining
the procedure described above for implementing $C_{jk}$, we note that a
two-targe-qubit CP gate $T_{jkl}$ described by Eq.~(3) can be realized using
four operational steps only:

Steps (i) and (ii): the operations for these two steps are the same as those
for steps (i) and (ii) described above.

Step\ (iii): Adjust the level spacings of qutrit $l$ such that the $%
\left\vert 1\right\rangle \leftrightarrow \left\vert 2\right\rangle $
transition is on resonance with the cavity. After an interaction time $%
t_{3}=\pi /g_{l12}$ (where $g_{l12}$ is the coupling constant between the
cavity mode and the $\left\vert 1\right\rangle \leftrightarrow \left\vert
2\right\rangle $ transition of qutrit $l$), the state $\left\vert
1\right\rangle _{l}\left\vert 1\right\rangle _{c}$ becomes $-\left\vert
1\right\rangle _{l}\left\vert 1\right\rangle _{c}$ while the states $%
\left\vert 0\right\rangle _{l}\left\vert 0\right\rangle _{c},$ $\left\vert
1\right\rangle _{l}\left\vert 0\right\rangle _{c}$ and $\left\vert
0\right\rangle _{l}\left\vert 1\right\rangle _{c}$ remain unchanged.

Step (iv): the operation for this step is the same as that for step (iii).

During performing single-qubit-gate operations, all three superconducting
phase qutrits $1,$ $2,$ and $3$ need to be decoupled from the cavity mode;
and during performing a two-qubit CP gate or a two-target-qubit CP gate,
irrelevant qutrits need to be decoupled from the cavity mode. This
requirement can be met by a prior adjustment of the level spacings of the
qutrits. Note that for superconducting qutrits, the level spacings can be
rapidly adjusted by varying external control parameters (e.g., magnetic flux
applied to phase, transmon, or flux qutrits; see, e.g., [3,36,37]).

As a final note, it should be mentioned that the method described above for
implementing \textit{a two-qubit CP gate} via resonant interaction is not
new, which was previously proposed [38,39]. We would like to stress that our
focus is to take the resonant interaction as an example to explore the
possibility of implementing the three-qubit DJ algorithm with
superconducting transmon qutrits coupled to a single cavity, by using the
protocol presented in the previous section.

\subsection{Fidelity}

\begin{figure}[tbp]
\includegraphics[bb=101 426 515 664, width=10.5 cm, clip]{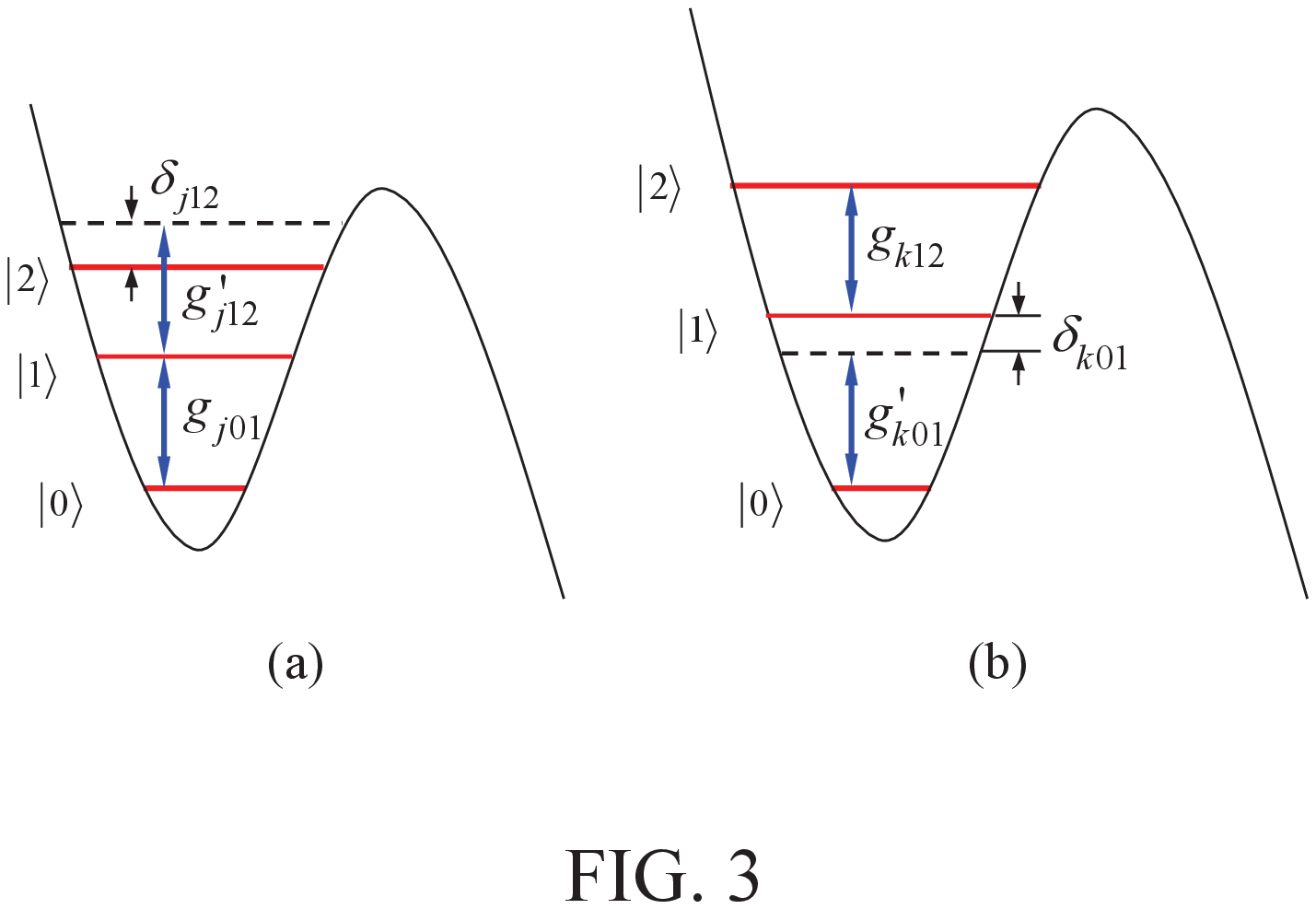} \vspace*{%
-0.08in}
\caption{(Color online) Illustration of qutrit-cavity interaction during the
cavity mode interacting with qutrit $j$ or qutrit $k$ ($j,k=1,2,$ or $3$).
For the details, see subsection 4.2.}
\label{fig:3}
\end{figure}

Let us now study the fidelity of the operation. Since the resonant
interaction is used in the implementation of the single-qubit $H$ gates or $%
\sigma _z$ gates, these basic gates can be completed within a very short
time (e.g., by increasing the pulse Rabi frequency). In addition, as
mentioned previously, one can apply the DRAG pulses to eliminate the leakage
into the level $\left| 2\right\rangle .$ Thus, the single-qubit gate error
is negligibly small. In this case, decoherence of the system would have a
negative impact on the operation of implementing a two-qubit CP gate as well
as a two-target-qubit CP gate,\textit{\ }due to the population of the cavity
photons during the operation. As discussed above, the implementation of
these CP gates involves two basic operations:

(i) The first one requires that during performing $C_{jk}$ and $T_{jkl},$
the cavity mode is resonant with the $\left| 0\right\rangle \rightarrow
\left| 1\right\rangle $ transition of the control qutrit $j.$ In realistic
case, the interaction Hamiltonian for this basic operation is given by
\begin{equation}
H_{I,1}=\left( g_{j01}a^{+}S_{j01}^{-}+h.c.\right) +\left( g_{j12}^{\prime
}e^{-i\delta _{j12}t}a^{+}S_{j12}^{-}+h.c.\right) ,
\end{equation}
where $a^{+}$ is the photon creation operator of the cavity mode, and the
second term represents the unwanted off-resonant coupling between the cavity
mode and the $\left| 1\right\rangle \leftrightarrow \left| 2\right\rangle $
transition, with a coupling constant $g_{j12}^{\prime }$ and a detuning $%
\delta _{j12}=\omega _{j12}-\omega _c<0$ [Fig.~3(a)].

(ii) The second one requires that during performing $C_{jk}$ and $T_{jkl},$
the cavity mode is resonant with the $\left| 1\right\rangle \leftrightarrow
\left| 2\right\rangle $ transition of the target qutrit $k.$ The interaction
Hamiltonian for this basic operation is given by
\begin{equation}
H_{I,2}=\left( g_{k12}a^{+}S_{k12}^{-}+h.c.\right) +\left( g_{k01}^{\prime
}e^{-i\delta _{k01}t}a^{+}S_{k01}^{-}+h.c.\right) ,
\end{equation}
where the second term represents the unwanted off-resonant coupling between
the cavity mode and the $\left| 0\right\rangle \leftrightarrow \left|
1\right\rangle $ transition, with a coupling constant $g_{k01}^{\prime }$
and a detuning $\delta _{k01}=\omega _{k01}-\omega _c>0$ [Fig.~3(b)].

As discussed previously, the cavity mode needs to be resonant with the $%
\left| 1\right\rangle \rightarrow \left| 2\right\rangle $ transition of the
target qutrit $l$ during performing $T_{jkl}.$ Note that the Hamiltonian
governing this basic operation is the same as $H_{I,2}$ with a replacement
of the index $k$ by $l.$

It should be mentioned that the term describing the pulse-induced or the
cavity-induced coherent $\left| 0\right\rangle \leftrightarrow \left|
2\right\rangle $ transition for each qutrit is not included in the
Hamiltonians $H_{I,1}$ and $H_{I,2}$, since this transition is negligible
because of $\omega ,\omega _c\ll \omega _{j02},\omega _{k02}$ ($j,k=1,2,3$)
(Fig.~3).

For each of the two basic types of operations described above, the dynamics
of the lossy system, composed of three qutrits ($1,2,3$) and the cavity, is
determined by
\begin{eqnarray}
\frac{d\rho }{dt} &=&-i\left[ H_I,\rho \right] +\kappa \mathcal{L}\left[ a%
\right]  \notag \\
&&+\sum_{j=1}^3\left\{ \gamma _{j21}\mathcal{L}\left[ S_{j21}^{-}\right]
+\gamma _{j20}\mathcal{L}\left[ S_{j20}^{-}\right] +\gamma _{j10}\mathcal{L}%
\left[ S_{j10}^{-}\right] \right\}  \notag \\
&&+\sum_{j=1}^3\gamma _{j,\varphi 2}\left( S_{j22}\rho S_{j22}-S_{j22}\rho
/2-\rho S_{j22}/2\right)  \notag \\
&&+\sum_{j=1}^3\gamma _{j,\varphi 1}\left( S_{j11}\rho S_{j11}-S_{j11}\rho
/2-\rho S_{j11}/2\right) ,
\end{eqnarray}
where $H_I$ is the $H_{I,1}$ or $H_{I,2}$ above, $j$ represents qutrit $j$ ($%
j=1,2,3$), $S_{j20}^{-}=\left| 0\right\rangle _j\left\langle 2\right| ,$ $%
S_{j22}=\left| 2\right\rangle _j\left\langle 2\right| ,$ $S_{j11}=\left|
1\right\rangle _j\left\langle 1\right| ,$ $\mathcal{L}\left[ a\right]
=\Lambda \rho \Lambda ^{+}-\Lambda ^{+}\Lambda \rho /2-\rho \Lambda
^{+}\Lambda /2$ with $\Lambda =a,S_{j21}^{-},S_{j20}^{-},S_{j10}^{-}$. In
addition, $\kappa $ is the decay rate of the cavity mode$,$ $\gamma _{j21,}$
$\gamma _{j20},$ and $\gamma _{10}$ are, respectively, the energy relaxation
rates of the level $\left| 2\right\rangle $ of qutrit $j$ for the decay
paths $\left| 2\right\rangle \rightarrow \left| 1\right\rangle $ , $\left|
2\right\rangle \rightarrow \left| 0\right\rangle $, and $\left|
1\right\rangle \rightarrow \left| 0\right\rangle ,$ and $\gamma _{j,\varphi
2}$ ($\gamma _{j,\varphi 1}$) is the dephasing rate of the level $\left|
2\right\rangle $ ($\left| 1\right\rangle $) of qutrit $j$.

The fidelity of the operation is given by [40]
\begin{equation}
\mathcal{F}=\sqrt{\left\langle \psi _{id}\right\vert \widetilde{\rho }%
\left\vert \psi _{id}\right\rangle },
\end{equation}%
where $\left\vert \psi _{id}\right\rangle $ is the output state of an ideal
system (i.e., without dissipation and dephasing) after a joint operation $%
U_{1},$ $U_{2},$ or $U_{3}$ is performed on the qutrit system initially in
the state $\left\vert 000\right\rangle $ and the cavity mode initially in
the vacuum state $\left\vert 0\right\rangle _{c}$, which is given by
\begin{eqnarray}
U_{1} &:&\left\vert \psi _{id}\right\rangle =\frac{1}{2}\left( -\left\vert
001\right\rangle +\left\vert 011\right\rangle +\left\vert 101\right\rangle
+\left\vert 111\right\rangle \right) \otimes \left\vert 0\right\rangle _{c}
\notag \\
U_{2} &:&\left\vert \psi _{id}\right\rangle =\frac{1}{2}\left( -\left\vert
001\right\rangle +\left\vert 011\right\rangle +\left\vert 100\right\rangle
+\left\vert 110\right\rangle \right) \otimes \left\vert 0\right\rangle _{c}
\notag \\
U_{3} &:&\left\vert \psi _{id}\right\rangle =\frac{1}{2}\left( -\left\vert
001\right\rangle +\left\vert 010\right\rangle +\left\vert 100\right\rangle
+\left\vert 111\right\rangle \right) \otimes \left\vert 0\right\rangle _{c},
\end{eqnarray}%
while $\widetilde{\rho }$ is the final density operator of the whole system
when the gate operations are performed in a realistic physical \textrm{system%
}.

\begin{figure}[tbp]
\includegraphics[bb=120 437 504 636, width=10.5 cm, clip]{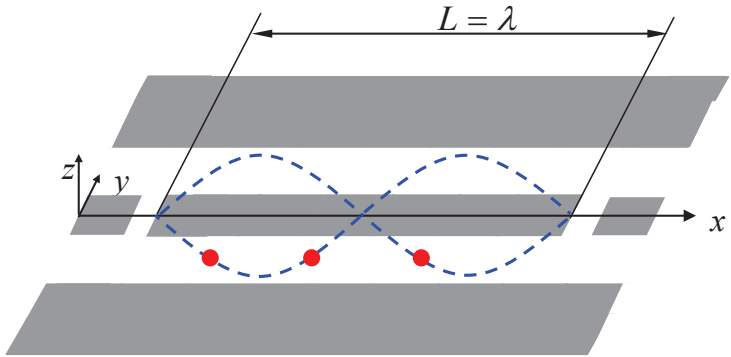} \vspace*{%
-0.08in}
\caption{(color online) Setup for three superconducting transmon qutrits
(red dots), and a (grey) standing-wave one-dimensional coplanar waveguide
resonator. $L$ is the length of the resonator, and $\protect\lambda$ is the
wavelength of the resonator mode. The two (blue) curved lines represent the
standing wave magnetic field in the $z$-direction.}
\label{fig:4}
\end{figure}

\begin{figure}[tbp]
\begin{center}
\includegraphics[bb=0 0 470 314, width=9.5 cm, clip]{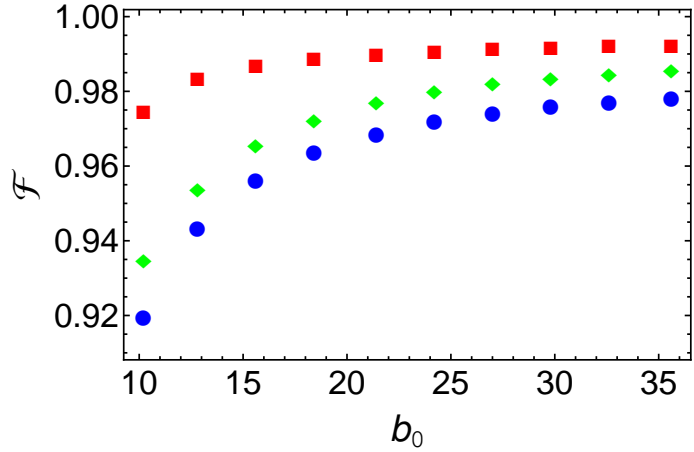} \vspace*{%
-0.08in}
\end{center}
\caption{(Color online) Fidelity versus $b_0$. Here, $b_0=\protect\delta %
_{01}/g_{01}^{\prime }$, and the red squares, green diamonds, and blue
circles correspond to the joint operations $U_{1},U_{2},$ and $U_{3}$ given
in Eq.~(4), respectively. Refer to the text for the parameters used in the
numerical calculation.}
\label{fig:5}
\end{figure}

We now numerically calculate the fidelity of the joint operations $U_{1,}$ $%
U_{2,}$ and $U_{3}$, for a setup shown in Fig.~4. Without loss of
generality, let us consider three identical transmon qutrits. In this case,
we can drop off the first subscript ($j,k,l$) for the detunings, Rabi
frequencies, and coupling constants. For simplicity, we assume that $%
g_{01}\sim g_{01}^{\prime }=g.$ One has $g_{12}\sim g_{12}^{\prime }\sim
\sqrt{2}g$ for the transmon qutrit here [41]. Choose $g/\left( 2\pi \right)
\sim 15$ MHz, which can be reached for a superconducting transmon qutrit
coupled to a one-dimensional standing-wave CPW (coplanar waveguide)
resonator [42]. Other parameters used in the numerical calculation are as
follows: $\gamma _{j,\varphi 2}^{-1}=\gamma _{j,\varphi 1}^{-1}=10$ $\mu $s,
$\gamma _{21}^{-1}=15$ $\mu $s, $\gamma _{20}^{-1}=150$ $\mu $s [43], $%
\gamma _{10}^{-1}=20$ $\mu $s, and $\kappa ^{-1}=5$ $\mu $s. Define $%
b_{0}=\delta _{01}/g_{01}^{\prime }$ and $b_{1}=-\delta _{12}/g_{12}^{\prime
}$. For simplicity, we choose $b_{1}=10.$ For the parameters chosen above,
the fidelity versus $b_{0}$ is shown in Fig.~5, from which one can see that
for $b_{0}=$ $24,$ a high fidelity $\sim 99.1\%,$ $98.0\%,$ and $97.2\%$ can
be achieved for the joint operations $U_{1},$ $U_{2},$ and $U_{3},$
respectively. We remark that the fidelity can be further increased by
improving system parameters.

For $b_{0}=24$ and $b_{1}=10,$ we have $-\delta _{12}\sim 0.21$ GHz, $\delta
_{01}\sim 0.36$ GHz, which is achieved in experiments [44]. $T_{1}$ and $%
T_{2}$ can be made to be on the order of $20-60$ $\mu $s for
state-of-the-art superconducting transom devices [4]. For superconducting
transmon qutrits, the typical transition frequency between two neighbor
levels is between 4 and 10 GHz [6,21,23,24]. As an example, let us consider
a cavity with frequency $\nu _{c}\sim 6$ GHz. Thus, for the $\kappa ^{-1}$
used in the numerical calculation, the required quality factor for the
cavity is $Q\sim 1.9\times 10^{5}$. Note that superconducting CPW resonators
with a loaded quality factor $Q\sim 10^{6}$ have been experimentally
demonstrated [45,46], and planar superconducting resonators with internal
quality factors above one million ($Q>10^{6}$) have also been reported
recently [47]. Our analysis given here demonstrates that implementation of
the three-qubit refined DJ algorithm is feasible within the present circuit
QED technique.

\section{CONCLUSION}

We have proposed a protocol for constructing the 35 $f$-controlled phase
gates of a three-qubit refined DJ algorithm, by using single-qubit $\sigma
_z $ gates, two-qubit CP gates and two-target-qubit CP gates. Using this
protocol, we have discussed how to implement the three-qubit refined DJ
algorithm with superconducting transmon qutrits resonantly coupled to a
cavity. Our numerical calculation shows that implementation of this quantum
algorithm is feasible for the current circuit QED. Finally, it is noted that
this protocol is quite general and can be applied to implement the
three-qubit refined Deutsch-Jozsa algorithm in various of physical systems.


\begin{thebibliography}{99}
\bibitem{s43} For a transmon qutrit with the three levels considered, the $%
\left\vert 0\right\rangle \leftrightarrow \left\vert 2\right\rangle $ dipole
matrix element is much smaller than that of the $\left\vert 0\right\rangle
\leftrightarrow \left\vert 1\right\rangle $ and $\left\vert 1\right\rangle
\leftrightarrow \left\vert 2\right\rangle $ transitions (see [38]). Thus, $%
\gamma _{20}^{-1}\gg \gamma _{10}^{-1},\gamma _{21}^{-1}.$
\end{thebibliography}

\section*{ACKNOWLEDGMENTS}

C.P.Y. was supported in part by the National Natural Science Foundation of
China under Grant Nos. 11074062 and 11374083, the Zhejiang Natural Science
Foundation under Grant No. LZ13A040002, and the funds from Hangzhou Normal
University under Grant Nos. HSQK0081 and PD13002004. Q.P.S. was supported by
the Zhejiang Provincial Natural Science Foundation of China under Grant No.
LQ12A05004. This work was also supported by the funds from Hangzhou City for
the Hangzhou-City Quantum Information and Quantum Optics Innovation Research
Team.

\end{document}